# Deterministic Thermal Sculpting of Large-Scale 2D Semiconductor Nanocircuits

*Maria Caterina Giordano, Giorgio Zambito, Matteo Gardella,*

*and Francesco Buatier de Mongeot\**

Dipartimento di Fisica, Università di Genova, Via Dodecaneso 33, 16146 - Genova, Italy

*E-mail: buatier@fisica.unige.it

**Abstract**

Two-dimensional (2D) Transition Metal Dichalcogenide semiconductor (TMDs) nanocircuits are deterministically engineered over large-scale substrates. The original approach combines large-area physical growth of 2D TMDs layer with high resolution thermal - Scanning Probe Lithography (t-SPL), to re-shape the ultra-thin semiconducting layers at the nanoscale level. We demonstrate the *additive* nanofabrication of few-layer $MoS_2$ nanostructures, grown in the 2H-semiconducting TMD phase, as shown by their Raman vibrational fingerprints and by their optoelectronic response. The electronic signatures of the $MoS_2$ nanostructures are locally identified by Kelvin probe force microscopy providing chemical and compositional contrast at the nanometer scale. Finally, the potential role of the 2D TMD nanocircuits as building blocks of deterministic 2D semiconducting interconnections is demonstrated by high-resolution local conductivity maps showing the competitive transport properties of these large-area nanolayers. This work thus provides a powerful approach to scalable nanofabrication of 2D nano-interconnects and van der Waals heterostructures, and to their integration in real-world ultra-compact electronic and photonic nanodevices.

## 1. Introduction

Emerging two-dimensional (2D) materials belonging to the class of Transition Metal Dichalcogenide semiconductors (TMDs) have recently gained a broad interest by the scientific community offering new routes to nanoscience and nanotechnology.[1–12] Thanks to their exceptional optoelectronic response and tunable bandgap in the Visible and Near-Infrared spectrum, combined with the atomically thin structure, functional properties can be obtained with a strong impact in various fields ranging from nanoelectronics and nanophotonics, to energy conversion and quantum technologies.[13–26] In this context the possibility to achieve a controlled reshaping of TMDs layers is particularly attracting in order to develop engineered and/or quantum confined 2D materials as a building block for functional nanodevices in electronics, photonics and quantum technologies[6,27–33]. Recently the possibility to promote tuning of the optoelectronic response via shape engineering of 2D layers has been demonstrated using micrometric TMDs flakes[34–36], highlighting the impact of this approach in photonics and quantum optics. However few attempts can be found on the arbitrary nanolithography of 2D TMD layers on large-scale wafers, typically applied to isolated micro-flakes[22,37,38], or characterized by spatial resolution at the micro-scale.[39]

So far, the most diffuse technique for 2D TMDs materials preparation has been mechanical exfoliation of crystals that provides randomly distributed 2D flakes, endowed with a size typically limited at the micro-scale. The urgent demand for potentially scalable platform and devices has recently motivated alternative growth methods for TMDs layers mainly relying on the Chemical Vapour Deposition (CVD) approach[40–47] which leads to a random distribution of triangularly shaped 2D TMD islands typically sized on a micrometer scale. Complex multi-step lithography processes, requiring chemical or plasma etching of the 2D islands (subtractive approaches), are typically used for nanopatterning 2D TMDs, introducing unwanted contaminations or surface damages to the fragile ultra-thin layers.[48,49] Scalable growth and non-invasive nanopatterning approaches are thus urgently required to engineer the optoelectronic and photonic response of 2D TMD layers.

The thermal-Scanning Probe Lithography (t-SPL) has recently emerged as a very promising technique, uniquely providing local modification of materials properties in ambient conditions with nanoscale spatial resolution provided by a sharp conductive probe[50,51]. This approach is optimal for the fragile 2D layers as demonstrated by few recent experiments showing the nanolithography of high quality metallic contacts on 2D TMDs[17], and their thermomechanical reshaping when exfoliated as micrometric flakes randomly distributed on the surface[52,53].

In this work we demonstrate a new *additive* approach enabling the scalable growth of ultra-thin 2D TMDs layer and their direct and high-resolution nano-sculpting via thermal-Scanning Probe Lithography. The nanolithography of ultra-thin $MoS_2$ nanocircuits deterministically located onto a large-scale wafer is demonstrated by combining the t-SPL of a sacrificial polymeric films, with large area ion beam assisted physical deposition of few-layer TMDs films. The Raman micro-spectroscopy maps have shown the characteristic vibrational response of the 2D semiconductor nanopaths, spatially engineered thanks to the t-SPL based method. The non-invasive t-SPL has been further exploited to precisely align simple nanodevices based on high-quality metallic contacts onto the 2D TMDs nanopaths, uniquely preserving their electronic response. The capability to control the electronic transport properties at nanometer lateral scale is demonstrated via high-resolution Kelvin probe nanoscopy and local probing of the electric transport via conductive-AFM nanoscopy. The local electrical and compositional contrasts and the electrical conductivity, resolved at the nanometer scale on the $MoS_2$ nanopaths, qualify them as building blocks of next generation nanocircuitry. The exceptional uniformity of the 2D TMDs layers over large-area ($cm^2$), combined with the non-invasive t-SPL nanolithography enables the precise nanofabrication of ultra-thin nanocircuits and van der Waals heterostructures nanodevices over large-scale wafers, opening new perspectives in electronics, photonics and quantum technologies.

## 2. Results and discussion

The homogeneous growth of ultra-thin $MoS_2$ layers over large-scale is demonstrated exploiting a new ion-beam assisted approach. Controlled deposition of ultra-thin semiconductor films is achieved via collimated ion beam sputtering (IBS) of a stoichiometric $MoS_2$ target, faced to the substrate (e.g. silica, silicon). Under this configuration large-area homogeneous films can be achieved on areas exceeding several $cm^2$, as highlighted by the sample picture of **Figure 1a**, which shows a few-layer $MoS_2$ film (thickness $\sim 6nm$) deposited on a transparent silica substrate. The deposition process of the pristine amorphous $MoS_2$ film takes place at room temperature and is thus compatible with flexible polymer substrates. Re-crystallization of $MoS_2$ films can be obtained via high temperature (750 °C) annealing in a tubular furnace, in presence of sulphur background pressure to avoid altering $MoS_2$ stoichiometry. The structural quality of the ultra-thin TMD film is confirmed by the Raman micro-spectra (**Figure 1b**) characterized by the $E^1_{2g}$ and by the $A_{1g}$ mode, resonant at 383 and 408 cm$^{-1}$ respectively, as expected for an ultra-thin $MoS_2$ layer [54]. Remarkably the Raman response, detected over sub-micrometric optical spots, is homogeneous up to the cm scale, as demonstrated by the spectra acquired along a diagonal axis of the sample, 5-6 mm apart one from the other (red spots in Figure 1a). The presence of a 2H-semiconducting $MoS_2$ ultra-thin film is further confirmed by the optoelectronic response detected in far-field extinction spectroscopy (**Figure S1**), and characterized by the A and B exciton resonances[55].

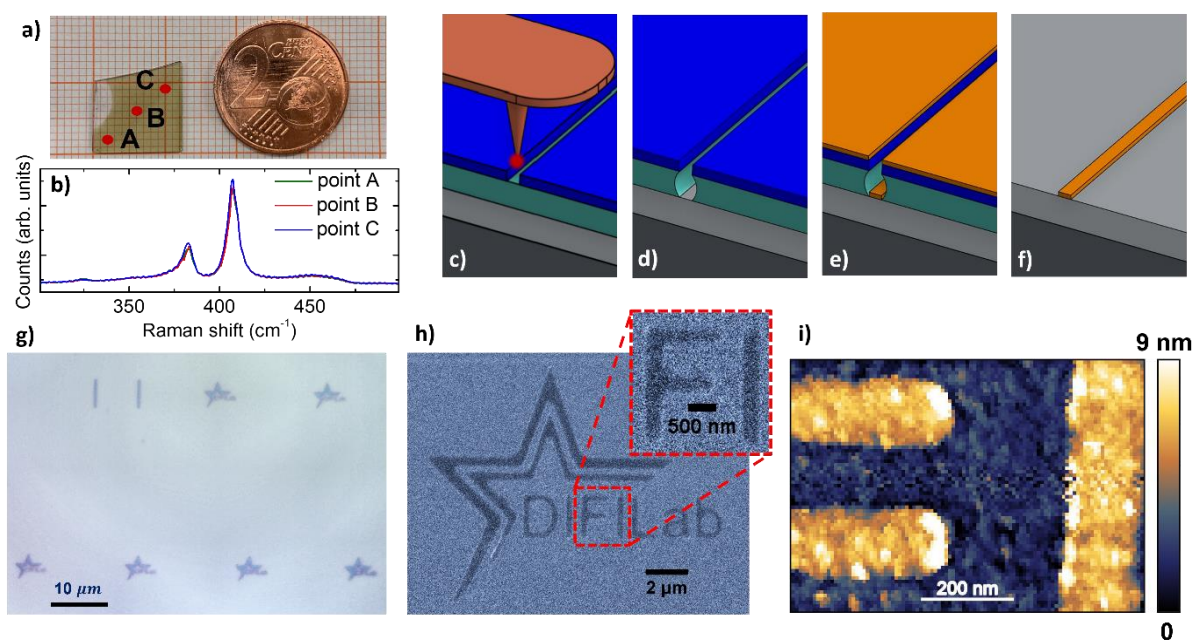

**Figure 1. a)** Picture of a large-scale ultra-thin $MoS_2$ layer grown onto a transparent silica substrate, and **b)** micro-Raman spectra detected in different spots of the samples shifted several mm apart one from the other (region A, B, C in the picture). **c-f)** Sketch of the t-SPL nanolithography process. **g)** Large-scale optical microscopy showing a series of $MoS_2$ nanocircuits fabricated onto a $Si/SiO_2$ substrate. **h)** SEM image of a few-layer $MoS_2$ nanocircuit (darker regions in the secondary electron signal) showing the logo of our laboratory (DIFILab facility - UNIGE) at the nanoscale, and zoom on a detail of the $MoS_2$ nanopaths. **i)** AFM image of a high resolution nanocircuit, showing ultra-thin $MoS_2$ nanopaths (scale bar 200 nm).

The large-scale homogeneity of these layers, combined with the novel t-SPL technique, provides the unique opportunity to develop a scalable approach for the nanoscale-resolved re-shaping of ultra-thin TMDs layers. The t-SPL indeed allows to pattern arbitrary nanopaths onto a thin polymeric bi-layer (see Methods) by exploiting a sharp hot nanoprobe (sketch in **Figure 1c**). In this way engineered nanopatterns can be *written* at high resolution (tip radius ~few nm) on the soft layer, which selectively exposes the substrate after due development process (**Figure 1d**). Hence the sacrificial bi-layer acts as a mask with appropriate negative angle profile for the following $MoS_2$ growth (**Figure 1e**). The final lift-off of the polymer film leaves behind engineered few-layer $MoS_2$ nanopaths, as sketched in **Figure 1f**.

The large-scale optical microscopy image of **Figure 1g** demonstrates the scalability of the method, enabling the *additive* nanofabrication of few-layer $MoS_2$ nanocircuits (darker regions of the image) in arbitrarily defined positions over large-scale, achieving uniform condition up to the cm$^2$ scale. The zoomed Scanning Electron Microscopy (SEM) image on a specific nanopattern (**Figure 1h**) shows well defined MoS$_2$ nanopath (darker contrast in the SEM image) forming the logo of our facility and obtained thanks to the t-SPL nanolithography. The zoomed-in SEM image well highlights the contrast of the few-layer MoS$_2$ on the substrate with nm spatial resolution. Further MoS$_2$ nanopatterns designed at higher resolution with a thickness of 6.3 nm ($\sim 9\ layers$) are shown in the AFM image of **Figure 1i**, which evidences nanostripe widths below 200 nm. Remarkably, the *additive* nanolithography approach here described enables fabrication of clean ultra-thin MoS$_2$ nanostructures, avoiding damaging and/or contamination of the 2D TMDs layer intrinsic in the reactive ion etching based approach, typically used for nanopatterning 2D materials.[48,56]

To confirm the material structure and to show the re-shaping capabilities of the few-layer material, Raman micro-spectra have been measured both on the engineered nanopaths (red curve in **Figure 2b** corresponding to blue regions of the optical microscopy image of **Figure 2a**) and few hundred nanometers apart on the bare substrate (black curve in Figure 2b). The Raman micro-spectrum detected on the MoS$_2$ nanopath shows the characteristic E$^1_{2g}$ and A$_{1g}$ vibrational modes excited at 383 and 408 cm$^{-1}$, respectively, while on silica substrate (black curve) we measure an unstructured background.

In parallel, the high homogeneity of the few-layers MoS$_2$ nanopaths is demonstrated by the micro-Raman maps shown in **Figure 2c** and **Figure 2d** respectively, corresponding to the Raman image of a whole nano-logo and of a zoomed detail highlighted in Figure 2a. These Raman maps, obtained at an excitation wavelength of 532 nm, recover the morphology of the few-layer MoS$_2$ nanostructures, with a spatial resolution in the range of few hundred nm's which is only limited by optical diffraction (Figure 2d).

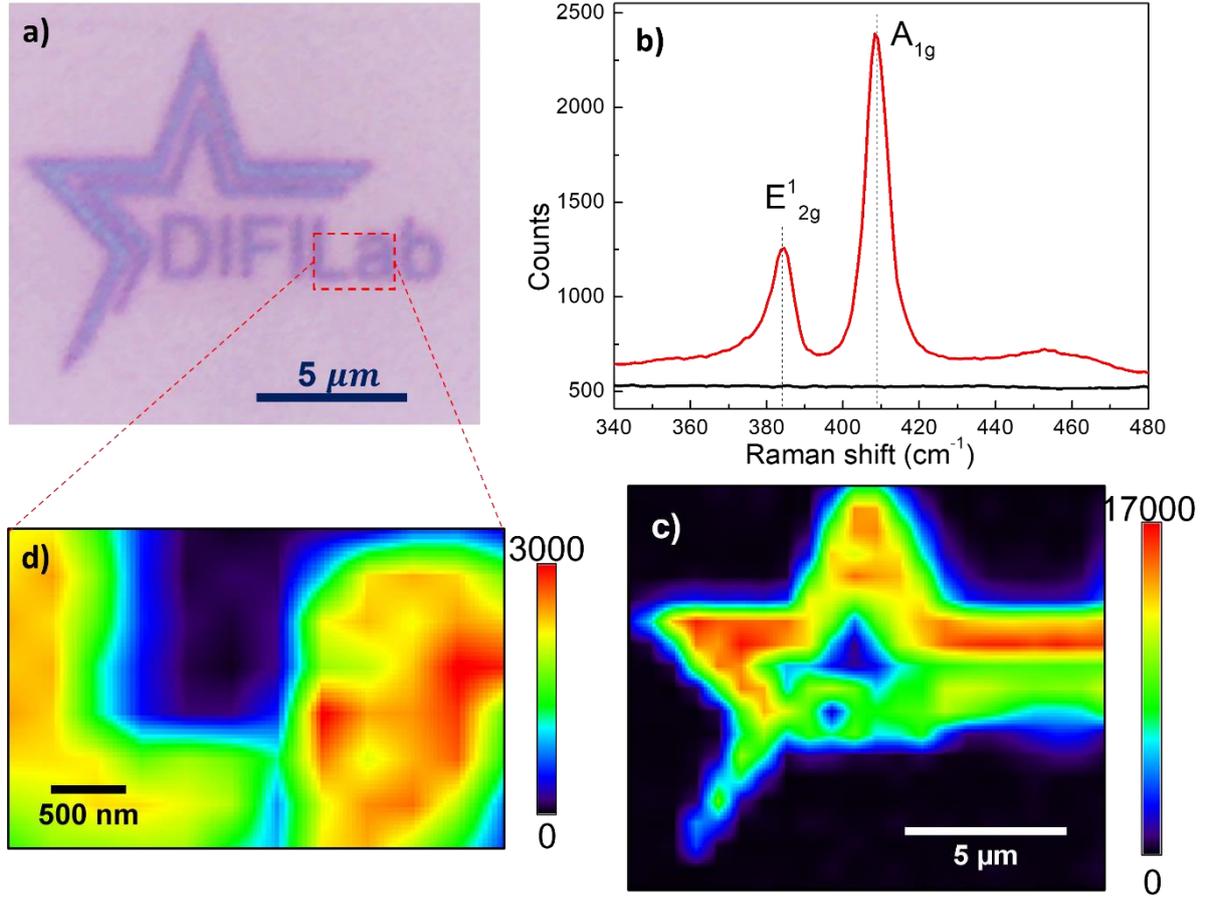

**Figure 2. a)** Optical microscopy image of a few-layer MoS$_2$ nanocircuit. **b)** Micro-Raman spectra detected on the MoS$_2$ nanopaths (red line) and on the neighboring Si region (black line). **c,d)** Micro-Raman maps detected on the whole MoS$_2$ nanocircuit and onto a detail (highlighted region in panel a), respectively. The maps show the signal detected in the range from 370 cm$^{-1}$ to 429 cm$^{-1}$, corresponding to E$^1_{2g}$ and A$_{1g}$ modes.

In order to reveal the spatial arrangement of the MoS$_2$ nanostructure, overcoming the optical diffraction limit, and to investigate the local electrical properties Kelvin Probe Force Microscopy (KPFM) was employed. Such analysis was carried out by using a Pt-coated conductive tip operating in single pass configuration, in order to extract contact potential difference maps of MoS$_2$ nanopaths lying on a Si/SiO$_2$ substrate. We define the contact potential difference (CPD) referred to SiO$_2$ work function as:

$$\Delta_{CPD} = \frac{1}{e}(\phi_{SiO_2} - \phi_s)$$

where e is the elementary electron charge, $\phi_{SiO2}$ is the silica work function and $\phi_s$ is the work function of the surface underneath the tip during the scan. **Figure 3a** shows the $\Delta_{CPD}$ map obtained on the MoS$_2$ nanopaths, highlighting their strong electrical contrast in terms of surface potential. An example of a $\Delta_{CPD}$ cross-section profile acquired across the MoS$_2$ nanotracks (white line in Figure 3a and corresponding profile in **Figure 3b**) shows a work function difference between MoS$_2$ nanopaths and silica substrate of about $200\ meV$. This electrical response is uniform over the CPD map, highlighting the MoS$_2$ to silica contrast with spatial resolution in the range of few tens of nm, determined by the Kelvin nanoprobe radius and by lock-in modulation voltage. In particular, the histogram of the $\Delta_{CPD}$ map (**Figure 3c**) is characterized by two distributions, respectively centered at 0 and -200 mV, as confirmed by the fit and by the cross-section profile at the MoS$_2$-silica edge. The former peak arises from the silica substrate and the latter from the MoS$_2$ nanopaths. The measured $\Delta_{CPD}$ map characterized by 200 mV contrast at the MoS$_2$-SiO$_2$ interface allows to quantify the work function of the MoS$_2$ nanopaths as $\phi_{MoS_2} \sim 5.25 eV$, calculated considering the value of SiO$_2$ work function ($\sim 5.05 eV$), in good agreement with respect to recent reports [57,58].

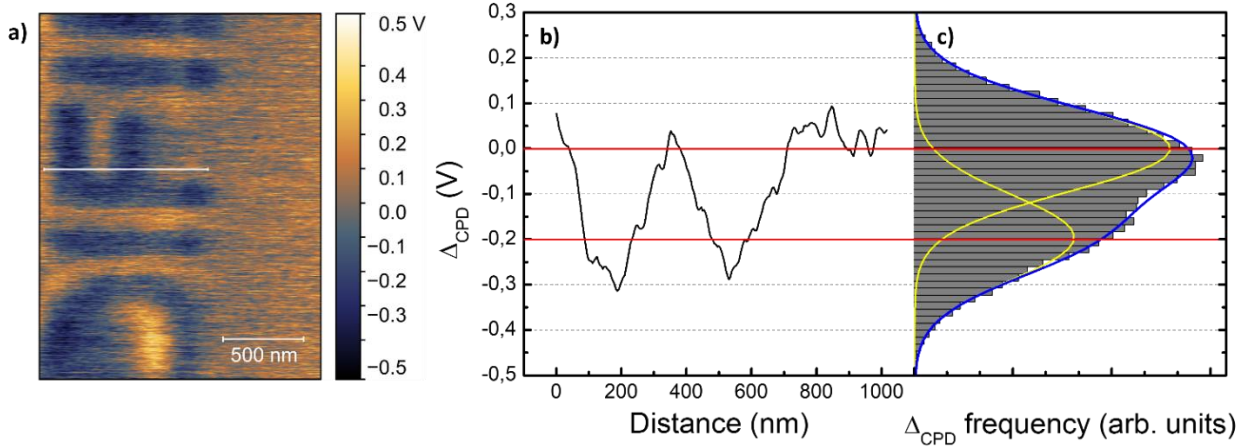

**Figure 3: a)** Kelvin Probe Microscopy image showing the Contact Potential Difference referred to silica work function $\phi_{SiO_2}$ ($\Delta_{CPD}$) signal, characterized by a strong electrical contrast of the MoS$_2$ nanocircuits. **b,c)** $\Delta_{CPD}$ line profile corresponding to white line in panel a), and histogram of the detected CPD signal, respectively.

In order to show the potential of the 2D-TMDs nanocircuits as active optoelectronic interconnects, we devise a simple ultra-thin device thanks to the non-invasive t-SPL approach. High quality metallic contact can be precisely aligned onto the nanocircuits by exploiting the peculiar in-situ imaging and real-time nanolithography capabilities (i.e. direct overlay method), avoiding undesired damage and contamination for the fragile layers[17]. In this way the electronic transport properties of the TMDs layers can be deeply investigated with nanoscale spatial resolution via conductive-AFM nanoscopy. **Figure 4** shows an example of 2D-TMDs device based onto a few-layer $MoS_2$ nanofinger. A metallic nanocontact (brighter finger at the bottom of the image) has been precisely aligned onto the few-layer nanofinger, thus enabling the local electric probing of the material with the conductive-AFM nanoprobe (ResiScope technology, sensitive across wide conductivity ranges). This c-AFM approach enabled the high-resolution detection of current and resistance maps onto the $MoS_2$ nanodevice, avoiding damage of the sample and/or the probe [59,60]. The c-AFM maps were acquired in contact-AFM configuration by applying a DC bias voltage of 0.5 V to the sample (metallic nanoelectrode) with respect to a p-doped single crystal diamond tip.

**Figure 4a** and **Figure 4b** show the local-current and -resistance map of the $MoS_2$ nanofinger device, detected in real-time with the topography (**Figure 4c**). A strong electrical contrast is detected at the $MoS_2$-substrate edges within both the current and the resistance map, as highlighted in **Figure 4d** by a comparison between topography and local current extracted along a horizontal profile (white dashed line in Figure 4a and Figure 4c). The electrical maps demonstrate the semiconducting behavior of the $MoS_2$ nanocircuit device with nanoscale spatial resolution, precisely corresponding to the AFM topography. The resistance (current) map shows a lower resistance (higher current) in proximity of the nanoelectrode, while a gradual resistance increase (current decrease) is detected as the electronic mean path to the contact increases. To quantify this effect, a current profile, extracted along a vertical profile of Figure 4a (dashed red line), is shown in **Figure 4e**. Due to the strong anisotropy of the $MoS_2$ nanofinger, endowed with nanoscale width, this curve well represents our system. The experimental data (black dots) are well described by a power law function $I(d) \propto d^{\alpha}$ with $\alpha = -1.1$ (red line in Figure 4e), where $d$ is the distance between the metal nanoelectrode and the conductive nanoprobe. This behavior well fits with the response of a semiconducting few-layer $MoS_2$ channel of length $d$ whose resistance $R_c$ is expected to scale as $R_c \propto \frac{1}{d}$. The local resistance maps allows to estimate a resistivity value of about 5 $\Omega m$ for these large-area few-layer $MoS_2$, under the approximation of a long channel device. This result is comparable to state of the art $MoS_2$ layers[33,61,62] and very promising in view of deterministic 2D TMDs nano-interconnects. Furthermore, these few-layer nanofinger configuration highlights the possibility to engineer ever more complex few-layer van der Waals semiconducting nanodevices, taking advantage of the non-invasive t-SPL nanolithography for the arbitrary and precise alignment of 2D materials endowed with their original optoelectronic properties.

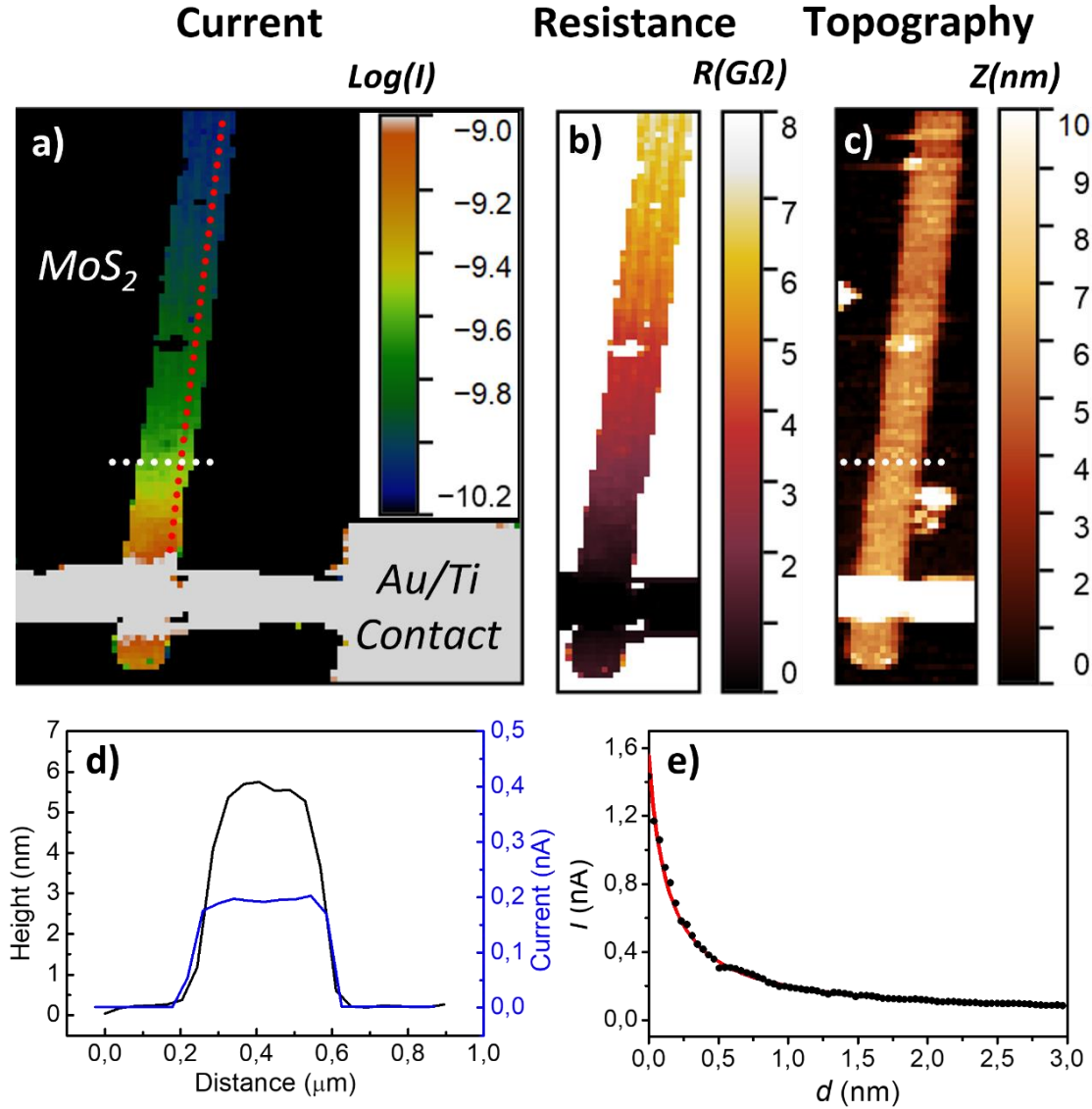

**Figure 4 a-c)** AFM current and corresponding local resistance and topography map of a few-layer MoS$_2$ nanostripe device under 0.5V DC bias. **d)** Current (blue) and corresponding topography (black) line profiles extracted across the substrate-finger edges (dashed white line in panel a and c). **e)** Current profile extracted from the current map of panel a) (dashed red line), showing a power law decay as the distance $d$ from the nanocontact increases. A power law $I(d) \propto d^{-1.1}$ (red line) is superimposed to the data (black dots).

## 3. Conclusion

We have shown the *additive* nanolithography of few-layer MoS$_2$ nanocircuits arbitrarily and precisely engineered onto large-scale wafers, thanks to the non-invasive thermal-Scanning

Probe Lithography approach combined with the homogeneous growth of few-layer TMDs across several cm$^2$. The thermal nanolithography of a sacrificial polymeric layer, combined with the large-scale physical deposition of the few-layer TMDs, uniquely enables the on-demand reshaping of ultra-thin TMDs layers with nanoscale spatial resolution. The so formed MoS$_2$ nanocircuits grow into the 2H-TMD semiconducting phase, as shown by their vibrational fingerprint in Raman micro-spectroscopy, and by their electrical work function in Kelvin probe nanoscopy. High quality few-layer TMDs nanodevices have been further engineered by exploiting the non-invasive t-SPL approach to precisely align metallic nanocontacts onto the MoS$_2$ nanocircuits. This configuration has enabled the local probing of the electronic transport properties of the few-layer MoS$_2$ nanocircuits, resolved via conductive-AFM nanoscopy. The local-conductivity maps highlight the competitive transport properties of these reshaped MoS$_2$ nanodevices as few-layer semiconducting interconnects in ultra-thin devices and components. Remarkably these results highlight the possibility to engineer even more complex van der Waals heterostructure nanodevices by this original t-SPL based technique, uniquely preserving the optoelectronic properties of the fragile few-layer materials. These reshaped 2D-TMDs nanocircuits thus open new perspectives for the integration of 2D semiconducting layers in scalable new-generation devices with impact in electronics, photonics, renewable energies and quantum technologies.

## 4. Methods

*Large area growth of 2D TMDs*

A fused silica (SiO$_2$) substrate is cleaned by ultrasound sonication both in acetone and isopropylic alcohol and loaded into a custom-made vacuum chamber with pressures of $10^{-6}$ / $10^{-7}$ mbar, faced towards a MoS$_2$ target. The latter is then irradiated by an ECR Plasma Source TPIS (TECTRA), which generates a 1.44 keV Ar+ ion beam (gas purity N5.0) at a pressure of $6.0 \cdot 10^{-4}$ mbar. The ion beam forms a 45° angle with respect to the MoS$_2$ target surface normal. The ion beam irradiation induces the sputtering deposition of MoS$_2$ thin films on the SiO$_2$ substrate, while the thickness of the deposited material is monitored by means of a calibrated quartz crystal microbalance.

A single zone tubular furnace is used for the recrystallization process (sulphurization). The sample is placed at the center of the furnace, while a 10 sccm flux of inert gas (Ar) is fluxed inside the pipe as a carrier gas. A quartz boat with sulfur powder is placed between the origin of the Ar flux and the sample. The furnace is then brought to the temperature of 750 °C with a heating ramp of 20°C/min and maintained at high temperature for a soaking time of 10 minutes. The heater is then turned off, allowing the cooldown of the sample.

*thermal-Scanning Probe Lithography*

The few layer $MoS_2$ nanocircuits are achieved thanks to the t-SPL patterning of a sacrificial bi-layer polymer mask, deposited by spin-coating onto a thermal oxide coated ($SiO_2$: 300 nm) p-doped silicon substrate. The underlying film is a PMMA/MA (thickness 95 nm) while a thermally sensitive film of polyphthalaldehyde (PPA - thickness 25 nm) is spin-coated on the surface. The new Nanofrazor Scholar setup (Heidelberg Instruments) enables the high-resolution t-SPL of the sensitive PPA film in ambient condition, thanks to the action of a sharp silicon tip (radius tip $\sim 10 nm$) locally heating the surface. The system is able to provide temperature- and time-controlled heat pulses through the nanoprobe (temperature at the cantilever sensor tunable in the 500 °C - 1100 °C range), thus enabling the arbitrary nanolithography of complex shapes and/or local modification of material properties.

A chemical development of the PMMA/MA film underlying the patterned nanopaths is finally performed in order to expose the patterned areas of the substrate (solution of deionized water in isopropylic alcohol at 5% vol.).

The sample with the patterned bi-layer polymer mask is transferred in the UHV system for the physical large-scale $MoS_2$ growth, as for the extended flat films (see "Large area growth of 2D TMDs" section). The sample is finally rinsed in acetone for the lift-off of the non-patterned areas achieving $MoS_2$ nanopaths on the substrate, avoiding damages and contaminations by the lithographic process. Under this condition high quality $MoS_2$ nanostructures can be easily achieved with the high-temperature recrystallization process.

Metallic nanocontacts based on a Ti/Au thin film (thicknesses 2 nm/20 nm) are precisely aligned onto the $MoS_2$ nanocircuits (see as a sake of example the few-layer $MoS_2$ nanofinger device of Figure 4) thanks to in-situ imaging and direct t-SPL overlay nanopatterning based on the polymeric bilayer process.

*Atomic Force Microscopy (AFM) characterization*

High resolution AFM images have been acquired by a JPK NanoWizard AFM (Bruker) operating in Quantitative Imaging (QI) mode.

The Kelvin Probe Force Microscopy (KPFM) maps have been acquired by a Nano Observer AFM (Concept Scientific Instruments) operating in single-pass mode and equipped with a platinum-coated silicon tip. The conductive AFM maps (current, resistance) have been acquired by using the same Nano Observer AFM, equipped with a ResiScope module enabling c-AFM nanoscopy over wide current (and resistance) ranges avoiding damage to the sample. The c-AFM maps have been detected with a conductive p-doped diamond tip scanning in contact mode.

The analysis of all the AFM images has been performed with the software Gwyddion (open source).

*Scanning Electron Microscopy (SEM) imaging*

The top-view SEM topographies of the $MoS_2$ nanocircuits (secondary electron signal) have been acquired by exploiting a thermoionic electron source biased at 5 kV (Hitachi SEM SU3500).

*Raman micro-spectroscopy*

Raman micro-spectra and maps have been detected by using an NRS-4100 Raman microscope (JASCO) operating in back-scattering configuration with sample excitation with a 532 nm laser source.

**Acknowledgements**

All the authors acknowledge R. Chittofrati, E. Vigo, F. Bisio, C. Canale and P. Canepa for technical assistance and useful discussions.

F. Buatier de Mongeot and M. C. Giordano acknowledge financial support by Ministero dell'Università e della Ricerca, within the project 'Dipartimento di Eccellenza 2018-2022' art. 1, c. 314-337, Legge 232/2016.

F.Buatier de Mongeot acknowledges financial support by Università degli Studi di Genova within the project BIPE 2020.

M. C. Giordano acknowledges financial support by Ministero degli Affari Esteri e della Cooperazione Internazionale (MAECI) within 'Progetti di Grande Rilevanza 2021-2023'- bilateral project Italy-Vietnam 'Large-area 2D/plasmonic heterostructures for photocatalysis and energy storage (H2D)'.

G. Zambito acknowledges support from Compagnia di San Paolo for financing his PhD scholarship.